\documentclass[]{aa}  
\usepackage{graphicx}
\usepackage{txfonts}
\usepackage{rotating}
\usepackage{natbib}
\usepackage{subfigure}
\bibpunct{(}{)}{;}{a}{}{,}

\def\msun{${\rm M}_{\odot}$}

\begin{document}

\title{The hydrodynamics of the supernova remnant Cas~A}
\subtitle{The influence of the progenitor evolution on the velocity structure and clumping
\thanks{See the online version for color figures.}
}

\author{B. van Veelen        \inst{1}
        \and
        N. Langer            \inst{1,2}
        \and
        J. Vink              \inst{1}
        \and
        G. Garc{\'i}a-Segura \inst{3}
        \and
        A.J. van Marle       \inst{4}
        }

\offprints{B. van Veelen}

\institute{  Astronomical Institute, Utrecht University, 
             P.O.Box 80\,000, 3508 TA Utrecht, The Netherlands\\
            \email{b.vanveelen@astro.uu.nl, n.langer@astro.uu.nl, j.vink@astro.uu.nl}
         \and
             Argenlander-Institut f$\mathrm{\ddot{u}}$r Astronomie, Universit$\mathrm{\ddot{a}}$t Bonn,
             Auf dem H$\mathrm{\ddot{u}}$gel 71, D-53121 Bonn, Germany\\
             \email{nlanger@astro.uni-bonn.de }
         \and
             Instituto de Astronom{\a'i}a-UNAM, 
             APDO Postal 877, Ensenada, 22800 Baja California, Mexico\\
            \email{ggs@astrosen.unam.mx}
         \and
             Centre for Plasma Astrophysics, KU Leuven,
             Celestijnenlaan 200B, bus 2400, B-3001 Leuven, Belgium\\
            \email{allardjan.vanmarle@wis.kuleuven.be}
           }
\date{ Received date /  Accepted date}

\abstract
{}
{There are large differences in the proposed progenitor models for the Cas~A SNR. One of these differences is the presence or absence of a Wolf-Rayet (WR) phase of the progenitor star. The mass loss history of the progenitor star strongly affects the shape of the Supernova remnant (SNR). In this paper we investigate whether the progenitor star of Cas~A had a WR phase or not and how long it may have lasted.
}
{We performed two-dimensional multi-species hydrodynamical simulations of the CSM around the progenitor star for several WR life times, each followed by the interaction of the supernova ejecta with the CSM. We then looked at the influence of the length of the WR phase and compared the results of the simulations with the observations of Cas~A.}
{The difference in the structure of the CSM, for models with different WR life times, has a strong impact on the resulting SNR. With an increasing WR life time the reverse shock velocity of the SNR decreases and the range of observed velocities in the shocked material increases. Furthermore, if a WR phase occurs, the remainders of the WR shell will be visible in the resulting SNR.}
{Comparing our results with the observations suggests that the progenitor star of Cas~A did not have a WR phase. We also find that the quasi-stationary flocculi (QSF) in Cas~A are not consistent with the clumps from a WR shell that have been shocked and accelerated by the interaction with the SN ejecta. We can also conclude that for a SN explosion taking place in a CSM that is shaped by the wind during a short ($\leq 15000~\rm yr$) WR phase, the clumps from the WR shell will be visible inside the SNR.}
{}


\maketitle

\section{Introduction}
\label{sec:intro}
One of the challenges of supernova remnant (SNR) research is to link the properties of the remnant with the properties and the evolution of the progenitor star. Ideally one would hope to learn about the late stages of stellar evolution from the properties of the SNR. In that respect one of the best-studied SNRs is the bright Galactic remnant Cas~A.

For a long time Cas~A has been thought to be the remnant of a very massive star, possibly that of an exploding Wolf-Rayet (WR) star \citep[e.g.][]{1987ApJ...313..378F,1993ApJ...411..823W}. In that case one would expect the supernova to be of Type Ib/c. However, recently \cite{2008Sci...320.1195K} obtained a spectrum of the supernova by observing its light echo, identifying it as a Type IIb explosion. This indicates that the progenitor lost most, but not all of its hydrogen envelope. 

X-ray observations indicate that the ejecta mass was only 2 - 4 \msun\ with about 1 - 2 \msun\ of oxygen \citep{1996A&A...307L..41V, 2003A&A...398.1021W}. The latter suggests an initial mass for the progenitor of approximately 20 \msun, which is close to the lower limit of the mass of a star at the start of the main sequence (MS) in the Galaxy that can become WR stars \citep{2005A&A...429..581M, 2000AJ....119.2214M}. However, it is not clear whether the low ejecta mass is due to heavy mass loss of a single massive star that was on its way, or just entered the WR phase, or whether mass loss was induced by a common envelope phase in a binary system \citep{2006ApJ...640..891Y}. Also the presence of a jet-counter-jet system in Cas~A \citep{2004NewAR..48...61V, 2004ApJ...615L.117H} puts strong constraints on the duration of a possible WR phase \citep{2008ApJ...686..399S}. Furthermore, the high density of the shocked circumstellar material (CSM) indicates that the shock wave is currently moving through the red super giant (RSG) wind of the progenitor \citep{2003ApJ...593L..23C}. 

Although recent studies \citep{2008ApJ...686..399S,2008Sci...320.1195K} seem to suggest an extended progenitor star, the presence of  a WR shell, i.e. the shell of red supergiant (RSG) wind material swept up by the fast WR wind, has been invoked to explain the presence of numerous slow moving \citep[$\leq 500$~km/s][]{1985ApJ...293..537V} nitrogen rich knots, often called quasi-stationary flocculi (QSF). \cite{1996A&A...316..133G} suggested that these knots are the remnants of the broken up WR shell. Note that also a binary common envelope phase is likely to be followed by a WR-like mass loss phase \citep{2006ApJ...640..891Y}. Both the fact that the QSF lie within the boundary of the forward shock, and the inferred density behind the forward shock of Cas A, limits the possible duration of a WR phase to $\sim 10^4$~yr \citep{1996A&A...316..133G}.

Here we report on our investigation of the imprint of the progenitor's stellar wind evolution, i.e. the existence and duration of a WR phase, on the morphology and kinematics of Cas A. For the first time we study the hydrodynamical evolution of a SNR in 2-D using multi-species advection, which allows us to separate the location and kinematics of wind material from the supernova ejecta.

\section{Method and Assumptions}
\label{sec:methodandassumptions}
\begin{table*}[htbp]
\centering
\caption{Adopted progenitor star wind parameters for our Cas~A hydrodynamic simulations: final evolutionary phase of the progenitor star, duration of this phase, corresponding mass loss rate and wind velocity, total amount of mass and kinetic energy lost during this phase, and model acronym.}
\renewcommand{\arraystretch}{1.1}
\begin{tabular}{c c c c c c c}
\hline
\hline
& $\tau [yr]$ & $\dot{M} [ $\msun$ /yr]$ & $v [km/s]$  & $\Delta M [$\msun$]$ & $ \Delta E [10^{45} erg]$ & Model\\
\hline
Red Supergiant & $8.75 \cdot 10^5$ & $1.54 \cdot 10^{-5}$ & $4.70$  & $13.5$ & $ 5.93$ & WR0\\
Wolf Rayet  & $5 \cdot 10^3$ & $9.72 \cdot 10^{-6}$ & $1.716 \cdot 10^3$  & $ 0.05$ & $1.42 \cdot 10^3$ & WR5\\
Wolf Rayet  & $10 \cdot 10^3$ & $9.72 \cdot 10^{-6}$ & $1.716 \cdot 10^3$  & $ 0.10$ & $2.85 \cdot 10^3$ & WR10\\
Wolf Rayet  & $15 \cdot 10^3$ & $9.72 \cdot 10^{-6}$ & $1.716 \cdot 10^3$  & $0.15$ & $4.27 \cdot 10^3$ & WR15\\
\hline
\end{tabular}
\label{table:CasA}
\end{table*}

\subsection{The adopted stellar evolution model}
\label{subsec:adopted}

As discussed in Section \ref{sec:intro}, the progenitor star of Cas~A has evolved through a RSG phase and perhaps a WR phase before exploding. This evolutionary history gives us constraints on the mass of the progenitor. Unfortunately, the allowed mass range of stars that have a WR phase during their evolution is not very well known. \cite{2005A&A...429..581M} show that stars with an initial mass as low as 20\msun\ may end up as a WR star, and that those stars will have a short WR life time. Alternatively, the progenitor of Cas A may have gone through a binary common envelope phase. However, we treat here the evolution of the progenitor as that of a single massive star. The reason for that is partly practical; there are no good analytical models of the hydrodynamics of a common envelope phase, and a full numerical simulation of this is beyond the scope of this study. Moreover, one may expect that a common envelope phase results in a non-spherically symmetric outflow. Although there is evidence for non-sphericity of the supernova ejecta in Cas A \citep[e.g.][]{2002A&A...381.1039W}, the outer shock wave is surprisingly circular \citep{2004NewAR..48...61V}, suggesting a more or less spherically symmetric CSM. The CSM structure that we assume, i.e. one that is shaped by a spherically symmetric wind of a single massive star, seems valid for Cas~A.

We have chosen an initial mass for the progenitor star of 20\msun, in accordance with previously suggested ZAMS masses \citep{2004NewAR..48...61V}. The stellar evolution model of this 20\msun, non-rotating star was calculated by  \cite{2004A&A...425..649H}. In this model the star does not have a WR phase and we thus assumed an enhanced RSG mass loss in such a way that during the RSG phase the entire hydrogen envelope is lost. It does not matter for the evolution of the core if the star loses its envelope entirely or only partly in this case, which means that we can assume the enhanced mass loss without affecting other parameters of the stellar evolution model. An enhanced mass loss rate is not unreasonable given by the uncertainty there currently is in RSG mass loss rates \citep[see][and discussion therein]{1988A&AS...72..259D,2005A&A...442..597V}.

The core mass at the end of the RSG phase is 6\msun which is also the final mass of the progenitor, due to our assumed enhanced mass loss rate. This mass is also roughly consistent with the ejecta mass of Cas~A of about 2 - 4\msun, taking the presence of the neutron star in Cas~A into account\citep{1999IAUC.7246....1T,2001ApJ...562..985C}. We adopt an ejecta mass of 4\msun, and an ejecta kinetic energy of $2 \cdot 10^{51}$ erg \citep{2004NewAR..48...61V}.

The mass loss rate and velocity of the stellar wind are calculated by combining the observational constraints and the parameters of the stellar evolution model. We can derive the total mass lost during the RSG phase by comparing the mass at the end of the MS phase, 19.5\msun, with the final mass of the progenitor, 6\msun. This gives a total mass loss of 13.5\msun. The RSG life time is $8.75 \cdot 10^5$ years which implies a mass loss rate of $1.54 \cdot 10^{-5}~\rm M_{\odot}~yr^{-1}$. Since we assume that all the mass is lost during the RSG phase, we also have to assume that during the  short duration WR phase the total amount of mass lost to the CSM does not increase significantly (see Table \ref{table:CasA}).

The radii of the forward and reverse shock of Cas~A are $2\farcm55 \pm 0\farcm2$ and $1\farcm58 \pm 0\farcm16$ \citep{2001ApJ...552L..39G}, which at a distance of $3.4_{-0.1}^{+0.3}$ kpc \citep{1995ApJ...440..706R}, correspond to 2.52 $\pm$ 0.2 pc and 1.58 $\pm$ 0.16 pc respectively. Since the amount of mass within the radius of the forward shock has to be $\sim$8\msun\ \citep{1996A&A...307L..41V}, we can determine the RSG terminal wind velocity by using the following relation between the mass loss rate, wind velocity and forward shock radius:

\begin{equation}
8 $\msun\ $ = \int \, 4 \pi r^2 \rho(r) \mbox{d}r = \int \, 4 \pi r^2 \frac{\dot{M}}{4 \pi r^2 \varv_{w}} \mbox{d}r = \frac{\dot{M} \cdot 2.52~\mathrm{pc}}{\varv_{w}}~,
\end{equation}
\noindent
in which we assume that the mass loss rate and the velocity are constant, and that the density is consistent with a $r^{-2}$ profile shaped by the free streaming RSG wind.

When we assume that the density profile, within the current forward shock radius, at the end of the RSG phase is consistent with an $r^{-2}$ profile shaped by the free streaming RSG wind, we have to check if the RSG shell is well outside the current shock radius.\footnote{This shell is created by the outflowing RSG wind and a hot MS bubble pushing inward.}  For this purpose we used equation (22) from \cite{1977ApJ...218..377W}, which gives the pressure inside the MS hot bubble, and the equation for the ram pressure of a stellar wind $P_{ram} = \rho v^2$. Using the parameters for the MS phase of the stellar wind given in \cite{2004A&A...425..649H}, a typical MS wind velocity of 1000 km/s, a ISM density of $10^{-23}~\mathrm{g}~\mathrm{cm}^{-3}$ and the parameters for the RSG wind given in Table \ref{table:CasA}, we obtain a radius for the RSG shell of approximately 4 pc. 

Because in our simulations we only consider the inner 3 pc of the CSM and the RSG shell is located at approximately 4 pc, the assumption of an initial $r^{-2}$ density profile at the end of the RSG phase for the inner 3 pc is justified. Since the structure of the CSM at the end of the RSG phase can be determined with the above mentioned assumptions, it was not necessary to perform a hydrodynamical simulation of the CSM during the RSG phase, which saved considerable computing time.

The values for the WR mass loss rate and wind velocity are calculated with help of the equations (12), (15), (17) and (22) from \cite{2000A&A...360..227N}. The parameters needed for the calculation of the mass loss rate and velocity of the WR wind are taken from the stellar model of \cite{2004A&A...425..649H}. All the values for stellar wind are summarized in Table \ref{table:CasA}. Since we do not know the WR life time, or whether there was a WR phase, we treat the WR life time as a free parameter. 

\subsection{Numerical method}
\label{subsec:code}

All the simulations presented here are done with the ZEUS MP code \citep{2006ApJS..165..188H}. This is a three dimensional Newtonian magneto-hydrodynamics code, which solves the Euler equations on a staggered mesh grid. Magnetism and gravity can also be treated by the code, but are not used here. 

In order to separate the CSM from the SN ejecta we use the multi-species advection available in ZEUS MP by considering two 'species', corresponding to the above mentioned components. The real composition of each of these species is not of importance since no feedback effects from the species back into the hydrodynamic calculations are taken into account.

Radiative cooling is included using the cooling curve from \cite{1981MNRAS.197..995M}, which is valid for a gas of approximately solar composition, but is applied to all the gas in the simulations. Applying this cooling curve to the gas consisting of CSM material is reasonable since it has a composition comparable to the solar composition. Applying it to the gas consisting mainly of SN ejecta material is not correct. However, the most important cooling in these regions is adiabatic cooling due to the expansion of the SN ejecta, and thus the error we make in this respect is small. By calculating how much energy we lose and assuming that all the gas is optically thin, we can use the cooling curve to estimate the radiative energy loss from these simulations.

Our simulations were done in two stages, the first stage was that of the interaction of the stellar wind with the CSM, which we calculated only once, and the second stage concerned the SNR evolution. Simulating the stellar wind was done in the same manner as described in \cite{1996A&A...316..133G}. In this method the innermost radial grid cells are given a density and velocity corresponding to the stellar wind at that point in time. This is done each time step to simulate the star blowing out its stellar wind. We simulated the evolution of the CSM due to the stellar wind only once, but used the output of the calculation at different evolutionary phases as input for stage 2, the collision of the supernova ejecta with the CSM. 

Simulating the supernova explosion was done in the same manner as described in \cite{2008ApJ...682...49W}. This method uses the following assumption for the radial density and velocity profile of the supernova ejecta, which at a given time has a flat inner core and a steeply declining outer edge:

\begin{equation} \label{eq:SNdens}
\rho(v,t) = \left\{ 
\begin{array}{lcr}
F \cdot t^{-3} & \mbox{for} & v \leq v_{core} \\
F \cdot t^{-3} \cdot (\frac{v} {v_{core}}) ^ {-n} & \mbox{for} & v_{core} < v \leq v_{max} \\
\rho_{CSM} & \mbox{for} & v > v_{max}
\end{array}
\right. ~,
\end{equation}

\begin{equation} \label{eq:freeexp}
v(r,t) = \frac{r}{t}~~\mathrm{for}~~t > 0 ~.
\end{equation}

\noindent Here $\rho$ is the density, $t$ is the time, $v$ is the velocity and $r$ is the radius. $F$ and $v_{core}$ are normalization constants which are determined through the assumed SN explosion energy and ejecta mass. The maximum velocity ($v_{max}$) is set to $3 \cdot 10^4$ km/s, which roughly corresponds to the maximum observed velocity in core collapse supernovae. The supernova explosion energy is assumed to be kinetic. The supernova ejecta are assumed to be spherically symmetric, so no clumps are present in the ejecta. The value for the exponent n is set to 9 for all simulations, which is usually assumed for core collapse supernovae \citep{1999ApJS..120..299T,2005ApJ...630..892D,2006ApJ...641.1051C}. We calculated four different SNR scenarios, corresponding to an explosion inside a CSM shaped only by the RSG wind (model WR0), and three explosions inside a CSM shaped by the WR star wind, with WR life times of 5000 yr (WR5), 10000 yr (WR10), and 15000 yr (WR15).

For all the results shown, a spherical coordinate system was used (r,$\theta$,$\varphi$), with an assumed symmetry in the $\varphi$-direction. The number of grid cell in the radial direction was always equal to 1000. The physical size, in the radial direction, of the grid in the pre-supernova evolution simulations was ranging from 0 to 3 pc, and the grid cells were equally spaced over this range. 

In the case of the supernova calculations the physical size of the radial grid was enlarged each time step, which is also done in the same manner as described in \cite{2008ApJ...682...49W}. For each time step the total number of grid cells was redivided over the grid, which resulted in an equally spaced grid for the radial direction at every time. The number of angular grid points, in the $\varphi$ direction, was equal to 200 for all simulations. The angular grid ranged from 0 to $\frac{\pi}{4}$ over which the number of grid cells were equally spaced.

Finally, we do not intend to match the observational properties of Cas~A to the last detail. This would require considerable fine tuning. For example, if we want to match our forward shock radius to the observations we could change both the SN explosion energy or the SN explosion mass which both affect the shock radius. We are more interested in looking at the influence of the presence or absence of a WR phase, and its duration, on the resulting SNR.

\section{Results}
\label{sec:results}
We have performed only one simulation for the evolution of the CSM prior to the supernova explosion. We take the output of this simulation at different times and use these different outputs as an input for the simulations of the SN ejecta interacting with the CSM.

\subsection{Pre-Supernova Evolution}
\label{subsec:PSN}

\begin{figure}
\centering
\includegraphics[width=0.60\columnwidth,angle=-90]{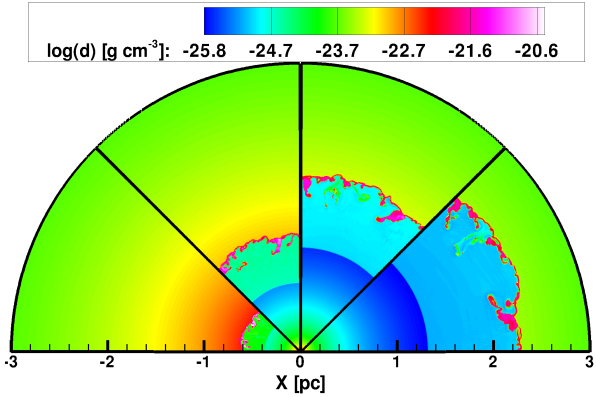}
\caption[CSM density structure evolution prior to the supernova explosion]
{From left to right the figure shows the CSM density structure approximately 5000, 10000, 15000 and 19000 years after the start of the WR phase. Looking from the first to the last slice (clockwise) the outward moving WR shell can be seen. This shell is unstable and will form clumps due to the Vishniac instability. Outside the WR shell the CSM structure is shaped by the wind of the previous RSG phase.
}
\label{fig:CasA-PSNda}
\end{figure}

\begin{figure}
\centering
\includegraphics[width=\columnwidth]{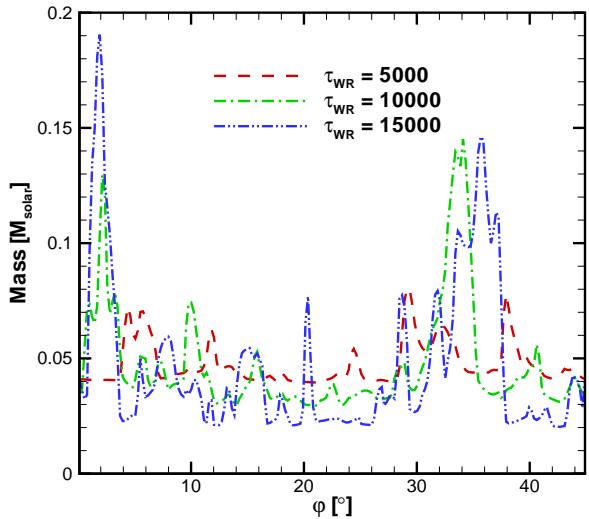}
\caption[]{Cumulative amount of mass in one angular grid cell and along all radial cells, from 0 to 3 pc, plotted against angle for Model WR5, WR10 and WR15. The zero degree angle corresponds to the left border of the slices of Fig. \ref{fig:CasA-PSNda}, with the 3 models corresponding to the first 3 slices. When the WR phase lasts longer with each model, the contrast of the total amount of mass at different angles also becomes larger. The peaks in the curves correspond to the angles at which clumps are located in the WR shell.
}
\label{fig:anglemass}
\end{figure}

\begin{figure}
\centering
\includegraphics[width=0.9\columnwidth]{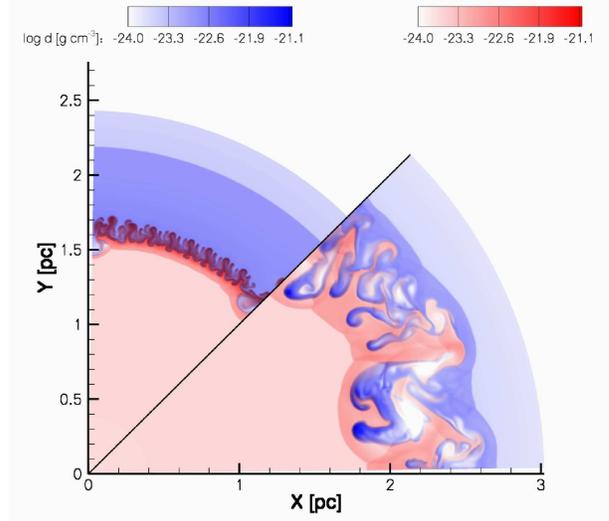}
\caption[]
{CSM (blue) and SN ejecta (red) density structure of the SNR of Models WR0 (first slice) and WR15 (second slice), 335 years after the supernova explosion. The blue color corresponds to the circumstellar material and red to the supernova ejecta. In regions where both material originating from the SN ejecta and from the CSM are present, the colors are added and will thus become purple. 
}
\label{fig:b36-d}
\end{figure}

Figure \ref{fig:CasA-PSNda} shows four snapshots of the CSM density structure approximately 5000, 10000, 15000, and 19000 years after the start of the WR phase. The figure shows that during the WR phase not a stable WR shell will form but a fragmented shell in which there are several small and larger clumps with a higher density than the density in the free streaming RSG wind. This clumping is due to the Vishniac/thin shell instability \citep{1983ApJ...274..152V,1993ApJ...407..207M,1995ApJ...455..160G}.

The clumping of the WR shell and its evolution affect the interaction of the supernova ejecta with the shell. We illustrate this in Fig. \ref{fig:anglemass} which shows the cumulative amount of mass for one angular grid cell and along all radial cells for the different angles of the three models with a WR phase. The relative difference between the amount of mass at angles where there is a clump compared to the angles where there is no clump becomes larger when the WR phase lasts longer. This implies that the clump masses grow with time. As a result of the clumping the supernova ejecta will be able to pass the WR shell easier at certain angles, which will affect the shock structure of the resulting SNR. When the WR phase lasts longer the clumps in the WR shell will be harder to destroy and to accelerate by the SN blast wave.

\subsection{Supernova Remnants}
\label{subsec:SN}

\subsubsection{SNR structure}
\label{subsubsec:clumps}

\begin{figure*}
\centering
\includegraphics[width=\textwidth]{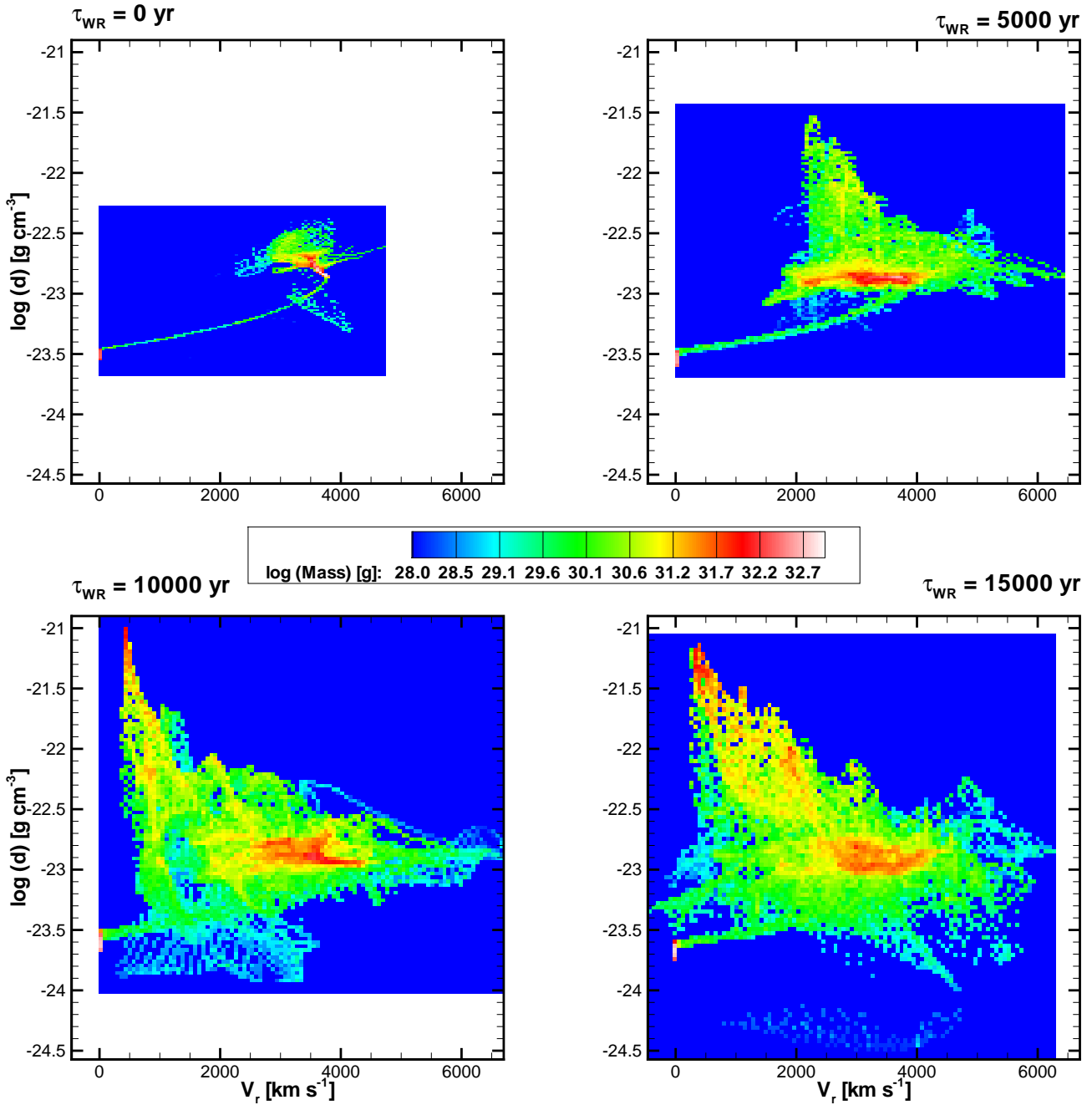}
\caption[]
{
Circumstellar mass histogram of Model WR0 (top left), WR5 (top right), WR10 (bottom left) and WR15 (bottom right), which shows the amount of circumstellar mass in a bin corresponding to a certain range in velocity and density. In all figures the high amount of mass at low density and low velocity represents the unshocked CSM. In the upper figures the curved line is an artifact from the artificial viscosity, which smears out the forward shock over several radial grid cells. For better comparison we have kept the color coding and the axis ranges the same for all subfigures.
}
\label{fig:CSM-hist}
\end{figure*}

Figure \ref{fig:b36-d} shows CSM and SN ejecta density structure of the resulting SNR for our models WR0 (first slice) and WR15 (second slice) 335 years after the supernova explosion. This corresponds approximately to the age of Cas~A assuming an explosion date of $1671 \pm 0.9$ \citep{2001AJ....122..297T}. In both slices the forward and reverse shock appear as a discontinuous jump in the density. The contact discontinuity (CD), which marks the boundary between the shocked SN ejecta and shocked CSM, can no longer be seen in the second slice, due to the collision of the SN ejecta with an irregularly shaped WR shell. In the first slice the fingers of the Rayleigh-Taylor instabilities at the CD show the mixing of the SN ejecta with the CSM. This instability is caused by the high density shocked SN ejecta being decelerated by the low density shocked CSM.

The most important difference between these two models is the different structure and corresponding density contrast within the SNR. Because the supernova ejecta encountered a smooth, spherically symmetric CSM in WR0 the only non-spherical component is the one caused by the Rayleigh-Taylor instability at the CD.  The large density contrasts within the second slice of Fig. \ref{fig:b36-d} are caused by the supernova ejecta colliding with a clumpy WR shell, instead of a smooth CSM. The high density clumps in the WR shell of the progenitor are not completely destroyed by the supernova ejecta and can still be seen as high density clumps within the SNR. Since they contain a lot of mass compared to their surroundings they are hard to accelerate and the supernova ejecta moved around them. This is reflected in the structure of the shocks, which is somewhat different for angles where there is a clump compared to the the angles where there is no clump.

\begin{figure*}
\centering
\includegraphics[width=\textwidth]{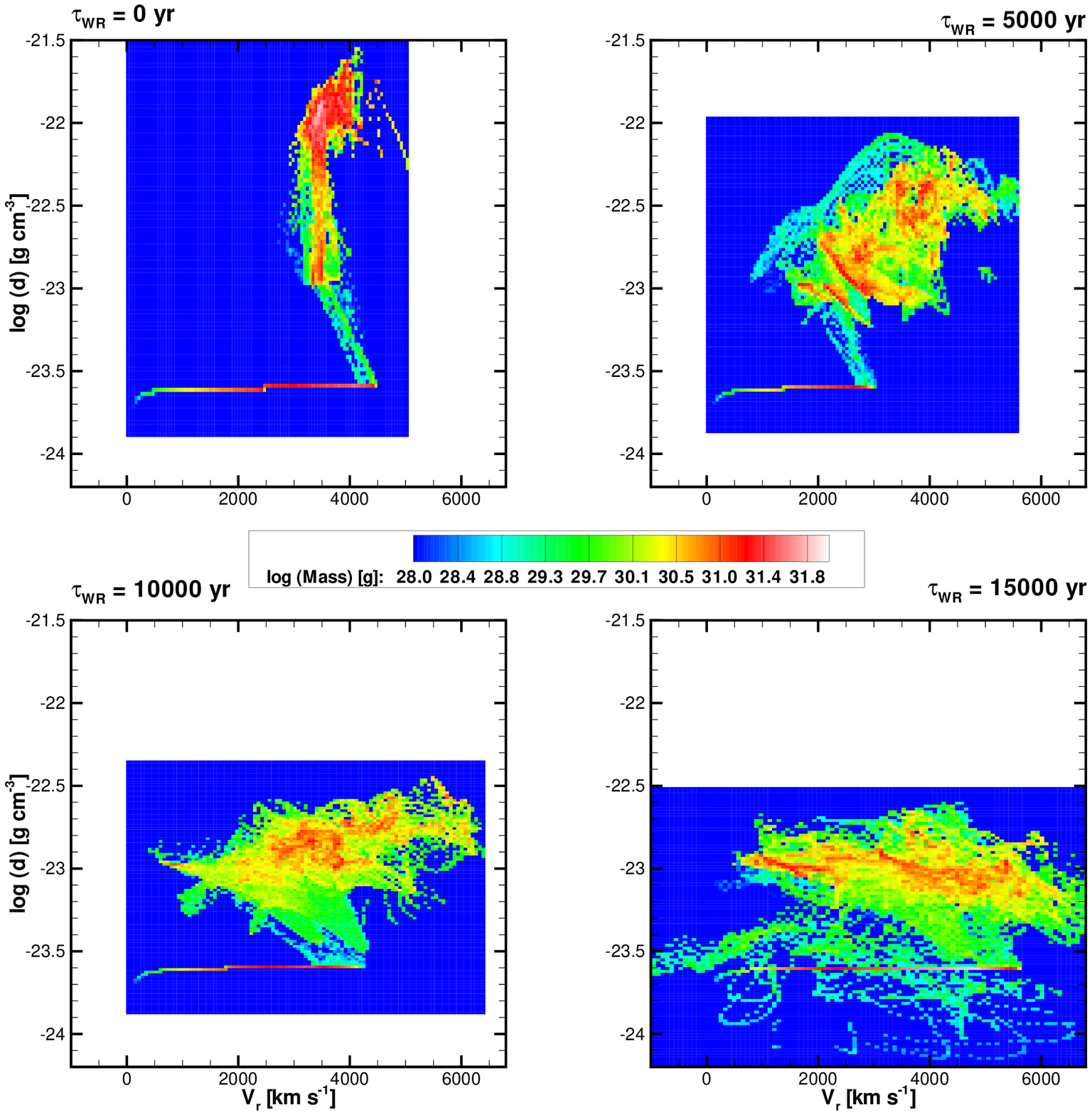}
\caption[]
{
Supernova ejecta mass histograms of Model WR0 (top left), WR5 (top right), WR10 (bottom left) and WR15 (bottom right), similar to Figure \ref{fig:CSM-hist}. The freely expanding ejecta can be seen as the horizontal line in all histograms. The rest of the mass in the histograms represents the shocked ejecta.
}
\label{fig:SN-hist}
\end{figure*}

\begin{figure}
\centering
\includegraphics[width=0.9\columnwidth]{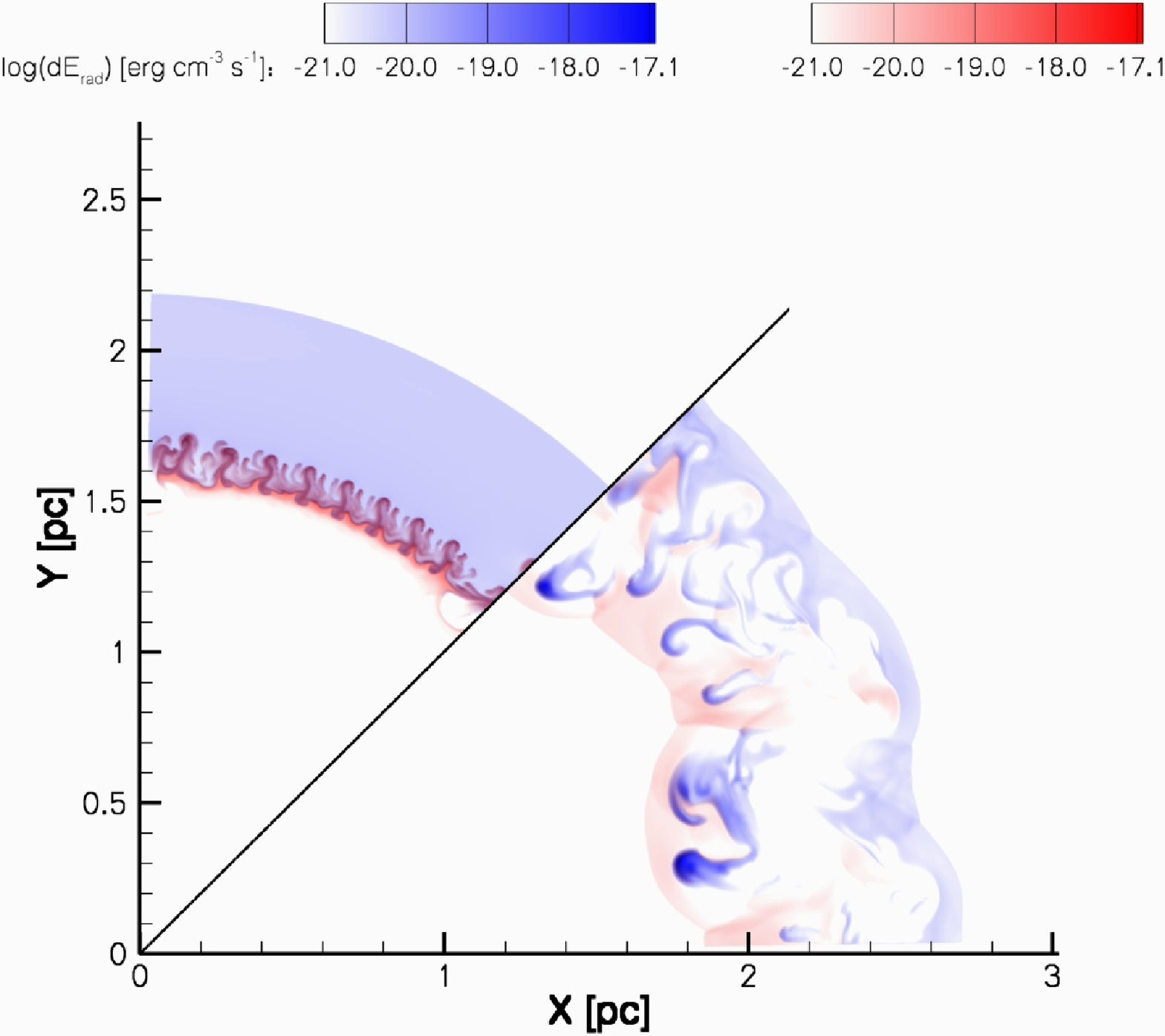}
\caption[]
{Radiative energy loss of the CSM (blue) and SN ejecta (red) components of the SNR Models WR0 (first slice) and WR15 (second slice), 335 years after the supernova explosion. The energy loss is calculated by using the \cite{1981MNRAS.197..995M} cooling curve. The second slice clearly shows that due to the clumped structure of the WR shell the majority of radiating material consists of shocked CSM. The fingers due to the Rayleigh-Taylor instability in the first slice show emission from both shocked CSM and shocked ejecta.
}
\label{fig:b36-dE}
\end{figure}

The distribution of CS mass with respect to the density and velocity can be seen in Fig. \ref{fig:CSM-hist}. These figures are mass histograms showing the amount of circumstellar mass in a bin corresponding to a specific range in velocity and density. They were made by adding the mass of every grid cell which had a CSM composition in excess of 70\%, placing it in the corresponding bin in velocity and density. In the upper left mass histogram of Fig. \ref{fig:CSM-hist} there are two regions with a considerable amount of CS mass. The first is the unshocked CSM which can be seen as the peak at low density and zero velocity. The second is the shocked CSM with a velocity exceeding 3500 km/s and a density of approximately $10^{-22.8}$ g$~\mathrm{cm}^{-3}$. All the other mass histograms still show the unshocked CSM as a high peak in mass at zero velocity and low density, but the shocked material is spread out over a much larger range in velocities, due to the violent interaction with the WR shell. It can also be seen that the range of velocities seen in the shocked CSM becomes larger with increasing WR life time. This is due to the increasing amount of mass within the WR shell for a longer WR life time. Another apparent component in the lower two histograms is the slow moving, highest density material which corresponds to the high density clumps in the remnant. Model WR5 does not show this feature but the remainders of the WR shell remain visible even in that model, as we will discuss below.

Figure \ref{fig:SN-hist} shows the mass histograms for the SN ejecta for all models. The unshocked, freely expanding ejecta are visible as a horizontal line in the histograms. Similar to what was seen in the shocked CSM distribution, the shocked SN ejecta of Model WR0 span a much smaller range in velocities compared to the shocked SN ejecta of the models with a WR phase. Also similar is that the range of velocities seen in the shocked SN ejecta increases with a longer WR life time.

Figure \ref{fig:b36-dE} shows the radiative energy loss of Models WR0 and WR15 corresponding to the density plots in Fig. \ref{fig:b36-d}. In the figure, only the material between the forward and reverse shock is visible. The material outside this region does not have a high enough temperature to emit enough radiation compared to the other components. In the first slice the Rayleigh-Taylor instabilities dominate, while in the second slice the high density clumps are most apparent. WR0 also shows that the RT instabilities are composed of both circumstellar material and SN ejecta. In the second slice the clumps are by far the most visible feature in the SNR, brighter by more than 2 orders of magnitude when compared to other material in the SNR. Although we believe that the qualitative result of the radiative energy loss is valid, one has to be cautious when looking at the results qualitatively. For instance, non equilibrium ionization and higher metal composition would increase line emission and are thus expected to increase the cooling rates.

\subsubsection{Shock structure and kinematics}
\label{subsubsec:shocks}

\begin{figure}
\centering
\includegraphics[width=\columnwidth]{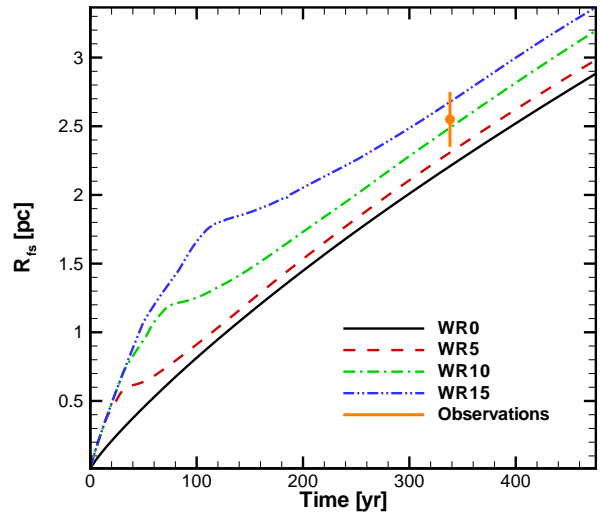}
\caption[]
{Angle averaged radius of the forward shock, for the different models, plotted against time. The orange dot shows the observational value for the forward shock radius at the age of Cas~A and the line shows the observational uncertainty in radius. The uncertainty in the age is too small to be visible in this figure. After the initial free expansion phase, the models with a WR phase collide with the WR shell at a different time and the expansion of the forward shock slows down. 
}
\label{fig:fs}
\end{figure}

\begin{figure}
\centering
\includegraphics[width=\columnwidth]{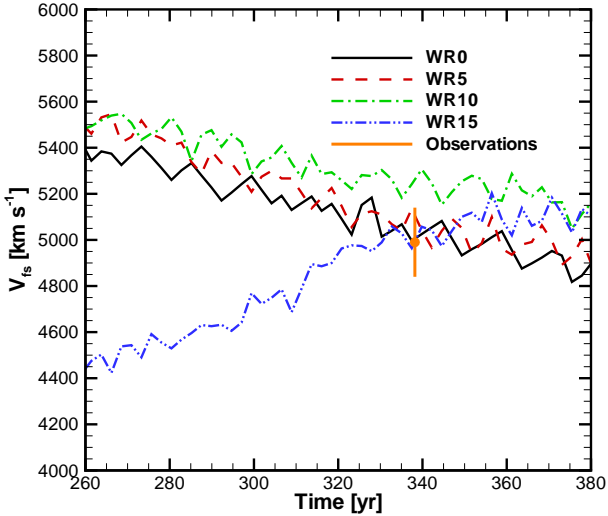}
\caption[]
{Velocity of the forward shock, for the different models, plotted against time. The observations and corresponding uncertainty are again shown in orange. Because of the large differences in the forward shock velocity during the total evolution of the SNR we only plotted a specific range in time. This range shows how the values for the forward shock velocity evolve around the age at which Cas~A is thought to be now. 
}
\label{fig:vfs}
\end{figure}

\begin{figure}
\centering
\includegraphics[width=\columnwidth]{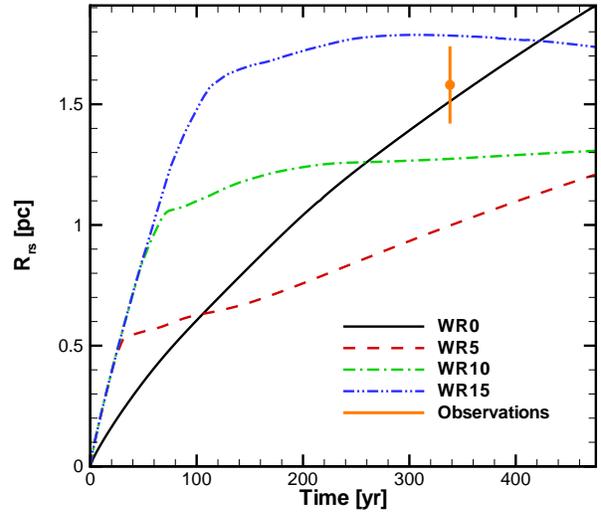}
\caption[]
{Similar to Figure \ref{fig:fs}, but in this case the reverse shock radius is plotted. In orange we show the observational value and its uncertainty. Initially all the reverse shock radii are the same for the models with a WR shell. However, the WR shell radius is different for each model and thus the onset of the decrease in expansion, i.e. the collision of the supernova ejecta with the WR shell, occurs at different times for WR5, WR10 and WR15, $\sim\!$ 30, $\sim\!$ 60 and $\sim\!$ 110 years respectively.
}
\label{fig:rs}
\end{figure}

\begin{figure}
\centering
\includegraphics[width=\columnwidth]{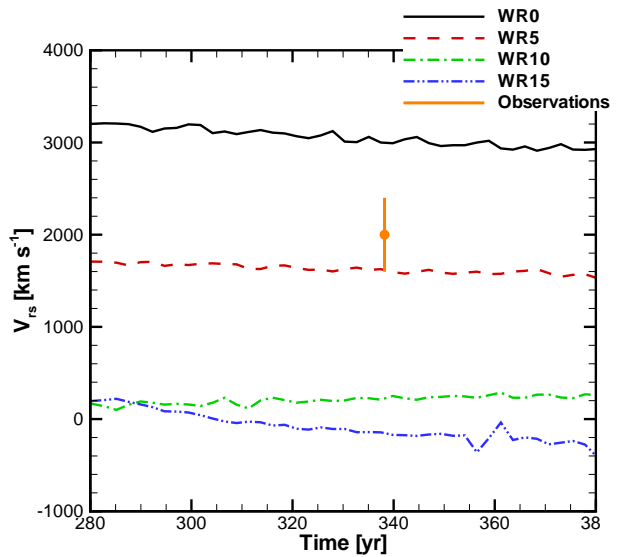}
\caption[]
{Similar to \ref{fig:vfs}, but in this case the reverse shock velocity is plotted. The models with a longer WR life time have a more massive WR shell which causes a higher inward velocity of the reverse shock. The model with the highest WR shell mass should thus show the largest inward reverse shock velocity in our frame, which is consistent with our results.
}
\label{fig:vrs}
\end{figure}

Figures \ref{fig:fs} to \ref{fig:vrs} show the angle-averaged values of the radius of the forward shock, the radius of the reverse shock, the velocity of the forward shock, the velocity of the reverse shock and the observed values. The radii of the forward and reverse shock were mentioned in Section \ref{subsec:adopted}, the velocity of the forward shock is $4990 \pm 150$ km/s \citep{1998A&A...339..201V,2004ApJ...613..343D,2008arXiv0808.0692P} and the velocity of the reverse shock is approximately $2000 \pm 400$ km/s \citep{2004ApJ...614..727M}. The value of the reverse shock velocity that is mentioned in the latter paper is actually 3000 km/s. However, that is measured in the frame of the unshocked ejecta, whereas we are considering the velocity of the reverse shock in the observers frame, and should thus use 2000 km/s, which also comes from that paper.

In Fig. \ref{fig:fs} it can be seen that for a given explosion energy and age of Cas~A, the longer the WR phase lasts, the larger is the radius of the forward shock. As the WR phase lasts longer, the low density region blown by the WR wind becomes larger, which enables the supernova ejecta to expand freely over a longer period which increases the forward shock radius. Although the measurement does not agree with the result of all models, the differences are small compared to the differences in the results for the reverse shock which we discuss below. A small change in the model parameters, or a difference due to the uncertainties within the models themselves, could already change the results such that all models could agree with the observations. The reason for the small variations in shock radius is that all models have roughly the same amount of swept up CSM. At the current age of Cas~A the forward shock velocities of all four models are also very similar to one another (Fig. \ref{fig:vfs}), and lie within 10\% of each other.

The comparison of the results for the reverse shock radius (Fig. \ref{fig:rs}) and the observations tell us more, since there is a clear difference between the models. For the models with a WR shell there are two important parameters which determine the radius of the reverse shock. The first is the radius at which the WR shell was located at the time of the supernova explosion, since that determines the time frame during which the supernova ejecta could expand freely. This is reflected in the different turn off times of the initial fast increase in the reverse shock radius for the models which contain a WR phase. Model WR0 does not expand as fast initially because the supernova ejecta encounter a higher density CSM, corresponding to the RSG wind, which creates a reverse shock that moves more slowly initially.

The second important parameter is the amount of mass in the WR shell compared to the supernova ejecta mass. Since the latter is constant, only the mass in the WR shell is important this case. The longer the WR phase lasts, the more mass is accumulated in the shell. As a result the collision between this shell and the supernova ejecta will be more violent and cause a reverse shock which moves inward faster with respect to the forward shock. This is represented in the time dependence of the reverse shock radii for the different models. The reverse shock radius of WR5 keeps on increasing after the collision with the WR shell while the reverse shock radius of WR15 starts moving inward after approximately 300 years.

This is also visible in Fig. \ref{fig:vrs}, which shows that the reverse shock velocity of WR15 becomes negative. By comparing the models in this figure we can see that for a longer WR life time, the velocity of the reverse shock at a given time will decrease, which is what we expect from the arguments given above. Together with the reverse shock radius, the velocity shown in Fig. \ref{fig:vrs} shows the largest differences between the four models. These differences give us the best tools to constrain the progenitor of Cas~A.

\section{Discussion and conclusions}
\label{sec:conclusions}
We have shown the results of 2-D simulations for the hydrodynamic evolution of a SNR in which, for the first time, the SN ejecta was separated from the CSM. With the Cas~A SNR in mind, we investigated the influence of the progenitor wind and the interaction of the supernova ejecta with the wind shaped CSM on the clumping and shock structure of the SNR.

From our simulations we can draw the following conclusions which can be compared with the observations of Cas~A:

\begin{itemize}
\item When a WR phase occurs, the SNR shows the remainders of the WR shell.
\item With an increasing WR life time, the reverse shock velocity in the observers frame, at the current age of Cas~A, decreases.
\item The longer the WR life time, the larger the range of velocities of both the shocked SN ejecta and the shocked CSM.
\end{itemize}

\begin{figure}[htbp]
\centering
\includegraphics[width=0.9\columnwidth]{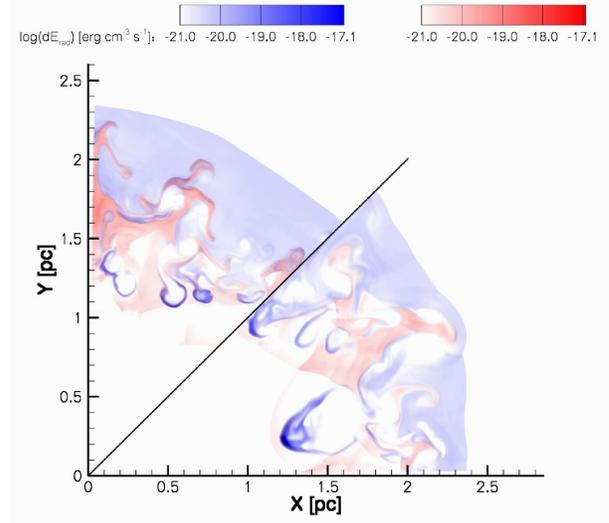}
\caption[]
{Radiative energy loss of the SNR of WR5 (first slice) and WR10 (second slice), similar to Figure \ref{fig:b36-dE}.  In both slices the clumps, which are the remnants of the WR shell, can be seen as the strongest emitting component.
}
\label{fig:b12-dE}
\end{figure}

The clumping that we find in the result of our SNR models that include a WR phase, is \emph{not} comparable to the QSF seen in Cas~A. In the highest density material we find a somewhat larger velocity range, extending to $\sim\!$ 1000 - $\sim\!$ 2000 km/s for Model WR10 and WR 15, depending on what density range one would consider as a clump. This is inconsistent with the observations for the QSF. For Model WR5 the clumps are destroyed, since no slow moving high density component is visible (Fig. \ref{fig:SN-hist}). Nevertheless, the remainders of those clumps should still be visible in the remnant and are even the dominant emitting component, which we illustrate by showing the radiative energy loss of WR5 and WR10 in Fig. \ref{fig:b12-dE}. This means that even for a very short WR life time of 5000 years remainders of a WR shell will be seen in the SNR. 

The clumps in our results are located in between the forward and the reverse shock, and their expansion in a 3D picture would correspond to a slowly expanding shell. Since all the emitting components of our result lie in between the two shocks, they should all show a similar distribution when projected onto the sky. This is in contrast with what is seen in the observations. Fig. 3 in \cite{1995AJ....109.2635L} shows that the QSF distribution does not coincide with the distribution of the other emitting components of the Cas~A SNR, and that the distribution of the QSF does not have a spherical shape. The same can be seen by comparing Fig. 5 and 10 in \cite{2001ApJS..133..161F}. This comparison also shows that there are some QSF outside the main shell in the southwestern part of the remnant, which does not occur in the models. Also, the spatial scale of the clumps in our results can range up to 0.2 pc, which is inconsistent with the measured sizes of the QSFs in Cas~A. The cooling times of the clumps in our results are of the order of hundreds of years ($\tau_{cooling} = \frac{e}{\dot{e}} \geq 100~\mathrm{yr}$), which does not correspond to a typical QSF life time of 25 years found by \cite{1985ApJ...293..537V}.

To be able to explain the QSF in Cas~A we would have to invoke another source of clumpiness, since our models are not able to explain the QSF. This other source would have to consist of clumps with properties similar to the QSF currently seen in Cas~A. The small size of the QSF in Cas~A tells us that they do not significantly influence the large scale dynamics of the SNR. The effect of the clumps from this other source would thus not affect the shock radii and velocities to a large extent and our conclusions would remain valid.

Due to our multi-species approach it can be clearly seen that clumps of SN ejecta can be found very close to the forward shock in models with a WR phase. This is solely due to the presence of a clumpy shell. Although it is of no specific significance to our current investigation, there are other remnants in which similar features are also observed and explained by invoking cosmic ray acceleration at the forward shock \citep{2005ApJ...634..376W,2008ApJ...680.1180C}. While our models and results do not apply to these specific SNRs, the fact that a clumpy CS shell can have this effect might be of interest to those investigations.

Cosmic ray acceleration is also invoked by \cite{2008arXiv0808.0692P}, who find that they need the energy loss due to this process to be able to explain the shock radii and velocities of Cas~A. We have not included this process in our simulations. Within our models there is a large set of parameters to be considered, and we have shown that within this parameter space it is possible to find radii and shock velocities which are largely consistent with the observed values. This does not mean we believe cosmic ray acceleration is not of importance, but given the uncertainty in the amount of energy involved in this process one can not unequivocally conclude that cosmic rays are needed in order to explain the structure and kinematics of Cas~A.

The differences in the forward shock radii and velocities of our results are not really meaningful, as we could accommodate those by small changes in explosion energy and ejected mass. However, the reverse shock structure varies considerably in our models, and can be used to constrain the progenitor model. The radii of the reverse shock in Models WR0 and WR15 lie closest to the observed value. From this figure alone one could argue that either a WR life time between 10000 and 15000 years or no WR phase at all would give us the required shock radius. By comparing the reverse shock velocity of WR10 and WR15 with the observations it can be seen that the difference is too large and that those models cannot explain the reverse shock velocity seen in the observations. The reverse shock velocity of WR0 does not match the observed value either, but the difference is smaller. If we want to match the reverse shock radius \emph{and} the reverse shock velocity at the same time, a model in which there is no WR phase seems most likely. However, recently there has been a debate about the reverse shock velocity in the western part of Cas~A, where the velocity is almost zero in our frame, and would be more in line with a long WR phase \citep{2008ApJ...686.1094H}.

Measurements of the reverse shock velocity are sparse and the value mentioned in \cite{2004ApJ...614..727M} is rather uncertain. Fig. 4.6 of \cite{2004DeLaneyThesis} shows that the proper motions of X-ray emitting components, which might \citep{2008ApJ...686.1094H}, or might not \citep{2008arXiv0808.0692P} be identified with material shocked at the reverse shock, show large differences in their expansion rates. In the west of the remnant the expansion rate seems to be smaller ($-0.1 \%\ \sim 0.1 \%\ \rm yr^{-1}$) than that of the majority of the material ($\sim 0.2 \%\ \rm yr^{-1}$). An expansion rate of $\sim 0.2 \%\ \rm yr^{-1}$ corresponds to a velocity of the X-ray emitting components of approximately 3000 km/s, given the reverse shock radius of 1.58 pc. This would imply a reverse shock velocity of approximately 2500 km/s, which is slightly higher than the value given in \cite{2004ApJ...614..727M} and would be even more consistent with Model WR0.

Our conclusion regarding the visibility of remainders of the WR shell in the SNR only holds for the short WR life times considered here. If the WR life time was much longer, the interaction of the WR wind with other parts of the CSM, the RSG shell for instance, would alter the structure of the CSM and consequently also the structure and visibility of the SNR. Nevertheless, this does not change our overall conclusion, since we can exclude longer WR life times on the basis of the shock structure and kinematics.

Because the presence of the QSF in Cas~A cannot be explained with help of the remainders of the WR shell and because the remainders of that shell do not correspond to any other observed component in the SNR, we can exclude the occurrence of a WR phase. This strengthens the growing evidence for not having a WR phase during Cas~A's progenitor life \citep{2008ApJ...686..399S,2008Sci...320.1195K,2009arXiv0905.1101P}.

\begin{acknowledgements}
We thank Michael L. Norman and the Laboratory for Computational Astrophysics for the use of ZEUS~MP. This work was sponsored by the Stichting Nationale Computerfaciliteiten (National Computing Facilities Foundation, NCF) for the use of supercomputer facilities, with financial support from the Nederlandse Organisatie voor Wetenschappelijk Onderzoek (Netherlands Organisation for Scientific Research, NWO).
\end{acknowledgements}

\bibliography{12393-1}
\bibliographystyle{aa}

\end{document}